\documentstyle[11pt,aaspp4]{article}
\input{psfig}

\received{}
\revised{}
\accepted{}
\journalid{}{}
\articleid{}{}
\paperid{}
\cpright{}{}
\ccc{}

\begin{document}

\title{Fluctuations in the Radio Background \\from Intergalactic
Synchrotron Emission}

\author{Eli Waxman\altaffilmark{1} and Abraham Loeb\altaffilmark{2}}
\altaffiltext{1}{Department of Condenced Matter Physics, Weizmann
Institute, Rehovot 76100, Israel; waxman@wicc.weizmann.ac.il}
\altaffiltext{2}{Harvard-Smithsonian Center for Astrophysics, 60 Garden
Street, Cambridge, MA 02138; aloeb@cfa.harvard.edu}

\begin{abstract}

The shocks produced in the intergalactic medium during large-scale
structure formation accelerate a population of highly relativistic
electrons which emit synchrotron radiation due to intergalactic magnetic
fields.  In a previous paper (Loeb \& Waxman 2000) we have shown that these
electrons cool primarily by inverse-Compton scattering of the microwave
background photons and can thereby produce the observed intensity and
spectrum of the diffuse $\gamma$-ray background.  Here we calculate the
intensity and angular fluctuations of the radio synchrotron background that
results from the same high-energy electrons, as well as the expected
angular fluctuations in the $\gamma$-ray background.  
On angular scales smaller than a degree, the predicted fluctuations in the
microwave background temperature are of order $40 \mu{\rm K}
(\xi_B/0.01)(\nu/10{\rm GHz})^{-3}$, where $\xi_B$ is the magnetic fraction
of the post-shock energy density.  This foreground might have already
dominated the anisotropy signal detected in existing low-frequency CMB
experiments, and can be identified with confidence through multi-frequency
observations.  Detection of the synchrotron fluctuations would allow to
determine the strength of the intergalactic magnetic field.  We predict a
strong correlation between high-resolution maps taken at low-frequency
radio waves and at high-energy $\gamma$-rays. Young X-ray clusters may also
appear as radio or $\gamma$-ray clusters. The detailed study of this
correlation will become easily accessible with the future launch of GLAST.

\end{abstract}


\section{Introduction}

More than a third of all X-ray clusters with luminosities $\ga 10^{45}~{\rm
erg~s^{-1}}$ possess diffuse radio halos (Giovannini, Tordi, \& Feretti
1999).  Based on energy arguments and circumstantial evidence, these radio
halos are believed to be caused by synchrotron emission from
shock-accelerated electrons during the merger events of their host clusters
(Harris et al. 1980; Tribble 1993; Feretti 2000; see Liang et al. 2000
for references to alternative, less successful models).
These highly-relativistic electrons cool primarily through inverse-Compton
scattering off the microwave background. Since their cooling time is much
shorter than the dynamical time of their host cluster, the radio emission
is expected to last as long as the shock persists and continues to
accelerate fresh electrons to relativistic energies. Intergalactic shocks
also occur along the filaments and sheets that channel mass into the
clusters. These structures, also traced by the galaxy distribution
(e.g. Doroshkevich et al. 1996), are induced by gravity and form due to
converging large-scale flows in the intergalactic medium.

In a previous paper (Loeb \& Waxman 2000), we have shown that most of the
diffuse $\gamma$--ray background (Sreekumar et al. 1998) might have been
generated by the shocks resulting from the formation of large-scale
structure in the intergalactic medium. These shocks produce a population of
highly-relativistic electrons with a maximum Lorentz factor $\ga 10^7$ that
scatter a small fraction of the microwave background photons in the
present-day universe up to $\gamma$-ray energies, thereby providing the
$\gamma$-ray background.  The predicted flux agrees with the observed
diffuse background over more than four decades in photon energy, provided
that the fraction of shock energy carried by relativistic electrons is
$\xi_e\sim0.05$ (a value consistent with that inferred for supernovae
shocks).  The brightest shocks are caused by the accretion onto (or mergers
of) X-ray clusters. The same electrons that emit $\gamma$-rays by
inverse-Compton scattering of microwave background photons, also produce
synchrotron radiation in the radio band due to intergalactic magnetic
fields. The existence of magnetic fields with an amplitude $\ga 0.1 \mu$G,
is inferred in cluster halos (Kim et al. 1989; Fusco-Femiano et al.  1999;
Rephaeli et al. 1999; Kaastra et al. 1999), and is also required for the
Fermi acceleration of these electrons.  The appearance of radio halos
around young X-ray clusters is therefore a natural consequence and an
important test of our model for the extragalactic $\gamma$-ray
background. The combination of radio and $\gamma$-ray data can be used to
calibrate $\xi_e$ and determine the strength of the intergalactic magnetic
field.

The intergalactic synchrotron emission contaminates the microwave
background anisotropies -- relic from the epoch of recombination, and as
such needs to be considered in the design and analysis of anisotropy
experiments at low frequencies.  Previous estimates of synchrotron
contamination of the cosmic anisotropies focused on Galactic emission,
which occurs primarily at low Galactic latitudes and large angular scales
(Tegmark et al. 2000).  In this {\it Letter}, we calculate the
intergalactic synchrotron contribution to the fluctuations in the radio sky
as a function of frequency and angular scale. We assume that most of the
emission originates from the virialization shocks around X-ray clusters,
and use the Press-Schechter mass function to describe the abundance of such
clusters as a function of redshift. Our simplified analytic calculation
should be regarded only as a first estimate of the synchrotron fluctuations
and could be refined with future numerical simulations (Keshet et
al. 2000).  Throughout the paper we adopt the popular set of cosmological
parameters (Ostriker \& Steinhardt 1995) with $\Omega_M=0.3$,
$\Omega_\Lambda=0.65$, $\Omega_b=0.05$, $h=0.7$, $\sigma_8=0.9$, and $n=1$.
In \S 2 we describe the method of the calculation. In \S 3 we discuss the
results and their main implications. Finally, we summarize our conclusions
in \S 4.

\section{Method of Calculation}

\subsection{Emission from a Single Halo}

Let us first consider the luminosity produced by inverse-Compton and
synchrotron emission of electrons accelerated to high energies in the
accretion shock around a halo of mass $M$ at a redshift $z$.  Assuming that
the shock is strong (i.e. the gas is not pre-heated), the population of
shock-accelerated electrons acquires a number versus energy distribution
with a power law index, $p\equiv d\ln N_e/d\ln E=-2$ (\cite{Bell78,BnO78}).
The inverse-Compton luminosity is then given by
\begin{equation}
\nu L_\nu^{\rm IC}= \frac{1}{2\ln(\gamma_{\rm max})}
\frac{\Omega_b}{\Omega_m} {\dot{M}(M,z)\over\mu m_p}
\xi_e \frac{3}{2} kT(M,z),
\label{eq:L_IC}
\end{equation}
where $\Omega_m=\Omega_M+\Omega_b$, $\mu m_p$ is the average particle mass
in the gas including electrons ($\mu=0.65$), $T$ is the post shock
temperature, $\gamma_{\rm max}\sim 10^7$ is the maximum Lorentz factor of
the accelerated electrons (Loeb \& Waxman 2000), $\dot M$ is the mass
accretion rate across the virialization shock, and $\xi_e$ is the fraction
of thermal shock energy carried by relativistic electrons.  The synchrotron
luminosity is then given by
\begin{equation}
\nu L_\nu^{\rm syn}= {u_B(M,z)\over u_{\rm CMB}}\nu L_\nu^{\rm IC},
\label{eq:L_syn}
\end{equation}
where $u_B$ and $u_{\rm CMB}$
are the energy densities of the post shock magnetic-field
and the cosmic microwave backround (CMB), respectively. 

The velocity 
dispersion, $\sigma$, of a halo can be related to its mass 
through the relation
\begin{equation}
M=\frac{\sqrt{2}}{5} {\sigma^3(M,z)\over GH(z)},
\label{eq:sigma}
\end{equation}
where 
\begin{equation}
H=H_0 a^{-3/2} g(a)
\label{eq:H}
\end{equation}
is the Hubble parameter, with $a\equiv(1+z)^{-1}$ and
$g(a)=[\Omega_m+\Omega_\Lambda a^3+(1-\Omega_m
-\Omega_\Lambda)a]^{1/2}$. The numerical constant was chosen so that, for
an isothermal sphere of velocity-dispersion $\sigma$, the quantity $M$ is
the mass enclosed within a sphere of average density $\bar\rho=200\rho_c$,
where $\rho_c=3H^2/8\pi G$ is the critical density at the corresponding
redshift. On dimensional grounds, the ensemble-averaged mass accretion rate
onto a halo of velocity dispersion $\sigma$ can be written as (White 1994)
\begin{equation}
\dot M(M,z)=f_{\rm acc}{\sigma^3(M,z)\over G},
\label{eq:M_dot}
\end{equation}
while the shock temperature and radius are given by 
\begin{equation}
kT(M,z)=f_T\mu m_p \sigma^2(M,z),
\label{eq:T}
\end{equation}
\begin{equation}
r_{\rm sh}(M,z)=f_{\rm sh}\frac{\sqrt{2}}{5} {\sigma(M,z)\over H(z)}.
\label{eq:r_sh}
\end{equation}
The factors $f_T$, $f_{\rm acc}$ and $f_{\rm sh}$ are dimensionless numbers
of order unity, which depend only weakly on the background cosmology for
the relevant range of $0.3 \la \Omega_m< 1$, and should be calibrated based
on numerical simulations (see, e.g. Miniati et al. 2000).  We have chosen
the normalization in equation (\ref{eq:r_sh}) so that for an isothermal
sphere density distribution, the post shock density is $\rho_{\rm
sh}=50\rho_c/3$, similar to the result for the self-similar accretion
solution in an Einstein de-Sitter ($\Omega_m=1$) cosmology (Bertschinger
1985). Note, that in this case the mass enclosed within $r<r_{\rm sh}$ is
$2M$.

We assume that the magnetic field energy density is a fixed fraction of the
post-shock energy density,
\begin{equation}
u_B(M,z)=\xi_B \frac{3}{2}kT(M,z)\frac{\Omega_b}{\Omega_m}
{\sigma^2\over2\pi G r_{\rm sh}^2\mu m_p}. 
\label{eq:u_B}
\end{equation}
With the above definitions, we find 
\begin{equation}
B(M,z)=0.14 \left[{\left(\frac{7\Omega_b}{\Omega_m}\right)
\left(\frac{\xi_B}{0.01}\right)f_T}\right]^{1/2}
f_{\rm sh}^{-1} h_{70}^{4/3}a^{-2}[g(a)]^{4/3}M_{14}^{1/3}\,\mu{\rm G},
\label{eq:B}
\end{equation}
where $H_0=70h_{70}{\rm km~s^{-1}\, Mpc^{-1}}$, $M=10^{14}M_{14}M_\odot$.
Thus, for values of $\xi_B\sim 0.01$ well below unity, we obtain a magnetic
field amplitude consistent with observations (Kim et al. 1989;
Fusco-Femiano et al.  1999; Rephaeli et al. 1999; Kaastra et al. 1999) and
with our model for the $\gamma$-ray background (Loeb \& Waxmann 2000).  The
implied synchrotron luminosity is
\begin{equation}
\nu L_\nu^{\rm syn}=5.1\times10^{38} f_{\rm acc}f_T^{-3/2}f_{\rm sh}^{-2}
\left({7\Omega_b\over\Omega_m}\right)^2 \left(\frac{\xi_B}{0.01}\right)
\left(\frac{\xi_e}{0.05}\right)
h_{70}^{2}a [g(a)]^{2}~T_{\rm keV}^{7/2}\,~~{\rm erg~s^{-1}},
\label{eq:L_syn_num}
\end{equation}
which can be related to the cluster mass through 
equation (\ref{eq:T}),
\begin{equation}
M_{14}=0.54 f_T^{-3/2} h_{70}^{-1}a^{3/2}[g(a)]^{-1}T_{\rm keV}^{3/2},
\label{eq:T_num}
\end{equation}
where $T_{\rm keV}=(kT/1{\rm keV})$.  
While $\gamma$-ray observations can be used to calibrate $\xi_e\sim 0.05$
(Loeb \& Waxman 2000), radio synchrotron observations of clusters calibrate
$\xi_B$.  The synchrotron luminosity predicted by equation
(\ref{eq:L_syn_num}) is consistent with the values observed for the
brightest radio halos of unrelaxed X-ray clusters for $\xi_B\sim
0.01$ (see the upper envelope in Fig. 9 of Liang et al. 2000, obtained for
$h=0.5$ and $z\sim 0.3$). Similarly to observations, our model predicts a
radio luminosity which increases with cluster temperature and a nearly flat
radio spectrum (see Fig. 6 in Liang et al. 2000).

The above expressions apply only to clusters which possess a strong shock
with a Mach number $\Upsilon\gg 1$. Non-accreting clusters or low-mass
clusters which accrete gas that was already pre-heated close to their
virial temperature (e.g., due to prior collapse of their environment or
because of supernova energy injection), will not show significant
non-thermal emission.  For a shock front moving at a moderate Mach number,
the accelerated electrons acquire a number-energy distribution with a
power-law index $p=(r+2)/(r-1)$, where $r=8\Upsilon^2/(6+2\Upsilon^2)$ is
the shock compression ratio for a gas with an adiabatic index $\gamma=5/3$.
This in turn yields steep non-thermal spectra with $\nu L_\nu^{\rm IC,syn}
\propto \nu^{-[2/(\Upsilon^2-1)]}$; implying that for weak shocks with
$\Upsilon\la 2$, most of the non-thermal emission is at undetectable low
frequencies.

\subsection{Background Flux and Anisotropy}

The observed intensity at Earth of inverse-Compton and synchrotron emission
is given by integrating (\ref{eq:L_IC}) and (\ref{eq:L_syn}) over
halo mass and cosmic time, $t$. Thus,
\begin{equation}
\langle\nu I_\nu^{\rm IC, syn}\rangle=\int {\rm d}z \frac{c {\rm d}t}{{\rm
d}z} \int {\rm d}M\, \frac{{\rm d} n}{{\rm d} M} \frac{\nu L_\nu^{\rm IC,
syn}(M,z)} {4\pi(1+z)^4},
\label{eq:I}
\end{equation}
where angular brackets denote sky average and ${{\rm d} n}/{{\rm d} M}$ is
the differential number density of halos per physical volume as a function
of their mass.  Using the Press-Schechter (1974) mass function, we find for
the cosmological parameters we have chosen, $\langle\nu I_\nu^{\rm
IC}\rangle =1.5f_{\rm acc}f_T(\xi_e/0.05){\rm keV~cm}^{-2}{\rm s}^{-1}{\rm
sr}^{-1}$. Thus, our model reproduces the $\gamma$-ray background intensity
for a reasonable choice of parameters, namely $f_{\rm
acc}f_T(\xi_e/0.05)\approx1$ (Loeb \& Waxman 2000).  Given the calibration
of the $\gamma$-ray background we predict a radio background with
\begin{equation}
\langle\nu I_\nu^{\rm syn}\rangle
=5.0\times10^{-12}f_{\rm acc}\left({f_T\over f_{\rm sh}}\right)^{2}
\left({\xi_e\over 0.05}\right)\left({\xi_B\over 0.01}\right)~~
{\rm erg}~{\rm cm}^{-2}~{\rm s}^{-1}~{\rm sr}^{-1}.
\end{equation}
We find that 90$\%$ (50$\%$) of the contribution to this background comes
from massive clusters with $M> 10^{14}M_\odot$ ($M> 5\times
10^{14}M_\odot$), which are less sensitive to pre-heating of the accreting
gas and for which the strong shock assumption is likely to apply.

The intensity fluctuations can be calculated from
\begin{eqnarray}
\delta^2 I(\theta)\equiv&&
\langle I_\nu(0)I_\nu(\theta)\rangle-\langle I_\nu\rangle^2\cr
=&&\int {\rm d}z \frac{c {\rm d}t}{{\rm d}z}
\int {\rm d}M\, \frac{{\rm d}n}{{\rm d} M} 
\frac{\left[{\nu L_\nu(M,z)}\right]^2}{16\pi^3(1+z)^8 r^2_{\rm sh}(M,z)}
P_{1,2}\left[{\theta d_A(z)\over r_{\rm sh}(M,z)}\right].
\label{eq:dI}
\end{eqnarray}
Here, $\langle I_\nu(0)I_\nu(\theta)\rangle$ is the sky average of the
product of intensities along two lines of sight, 1 and 2, separated by an
angle $\theta$, $d_A$ is the angular diameter distance, and $P_{1,2}$ is
the probability that line-of-sight 2 passes through a halo of radius
$r_{\rm sh}$ given that line-of-sight 1 passes through the same halo,
\begin{equation}
P_{1,2}(x)={2\over\pi}\int_0^\pi{\rm d}\phi\int_0^1{\rm d}y y
\times \cases{1,&for $(y+x\cos\phi)^2+x^2\sin^2\phi<1$;\cr 0,& otherwise
.\cr}
\label{eq:P12}
\end{equation}
We have neglected in equation (\ref{eq:dI}) halo-halo correlations, and
assumed that the only contribution to the integral is from both
lines-of-sights passing through the same halo. This assumption is justified
since the optical depth (the average number of halos along a given line of
sight) is much smaller than unity for the typical parameter values.  In
particular, we find that clusters with $M> 5\times 10^{14}M_\odot$
($M>10^{15}M_\odot$) cover $<10\%$ ($<2\%$) of the sky.  Note that the
fractional fluctuations, $\delta^2I/\langle I\rangle^2$ are independent of
$f_T$, $f_{\rm acc}$, $\xi_e$, and $\xi_B$, and their dependence on $f_{\rm
sh}$ is simply given by $f_{\rm sh}^{2}\delta^2I(f_{\rm sh}\theta)/\langle
I\rangle^2= [\delta^2I(\theta)/\langle I\rangle^2]_{f_{\rm sh}=1}$.

\section{Results}

The fluctuations in the radio and $\gamma$-ray background intensities,
derived from equation (\ref{eq:dI}), are shown in Figure 1.  Although the
synchrotron background amounts to only a small fraction of the CMB
intensity, $\langle I^{\rm syn}_\nu\rangle/I^{\rm CMB}_\nu=6\times10^{-6}
(f_T/f_{\rm sh}^2)(\xi_B/0.01)(\nu/{10\rm GHz})^{-3}$ for $f_{\rm
acc}f_T(\xi_e/0.05)=1$, its fluctuations could dominate over the primordial
CMB fluctuations at low photon frequencies, $\nu \la 10$~GHz.  Our results
imply that radio emission from cluster shocks contributes a fluctuation
amplitude between $\sim 30$ and $50 \mu$K $\times (f_T/f_{\rm
sh}^2)(\xi_B/0.01) (\nu/10{\rm GHz})^{-3}$ to the CMB on angular scales
between 1 and $0.1^\circ$, respectively.  
Interestingly, current anisotropy experiments are just sensitive to this
level of fluctuations\footnote{See table summary of current experiments at
http://www.hep.upenn.edu/$\sim$max/index.html, or at
http://cfa-www.harvard.edu/$\sim$mwhite/cmbexptlist.html}.  Existing
detections by CAT ($50\pm15\mu$K at 15~GHz on 0.2--0.5\arcdeg scales) and
OVRO ($56^{+8.5}_{-6.6}\mu$K at 20~GHz on 0.1--0.6\arcdeg scales), as well
as 95\% upper limits ($\la 40\mu$K on arcminute scales at 9-15 GHz by the
ATCA, RYLE and VLA detectors) are consistent with our prediction\footnote{
Note that although the cluster contribution declines rapidly at high
frequencies, some of these experiments, like CAT or OVRO, measured
fluctuations at only one frequency and could not reject a synchrotron
contribution to the measured signal.}.

For bright clusters, the flux increment due to synchrotron emission by the
non-thermal electrons accelerated in the cluster accretion shock is larger
at low frequencies than the Sunyaev-Zeldovich decrement due to the thermal
intra-cluster electrons. For a gas density profile of an isothermal sphere,
\begin{equation}
{\Delta I_\nu^{\rm syn}\over \Delta I_\nu^{\rm SZ}} =-3.8 f_{\rm acc}f_T
f_{\rm sh}^{-3} \left({7\Omega_b\over\Omega_m}\right)
\left(\frac{\xi_B}{0.01}\right) \left(\frac{\xi_e}{0.05}\right)
h_{70}^{11/3}a^{5/2}[g(a)]^{11/3}M_{14}^{2/3}\left({\theta d_A\over r_{\rm
sh}}\right) \left({\nu\over {10\rm GHz}}\right)^{-3}.
\label{eq:SZ}
\end{equation}
In the X-ray regime, the central surface brightness due to thermal
bremsstrahlung emission by the intra-cluster gas exceeds by orders of
magnitude the inverse-Compton brightness; however, due to the rapid decline
of the thermal brightness with projected radius outside the cluster core
($\propto \theta^{-3}$), we find that the non-thermal brightness might
dominate around the shock radius.

The cumulative (all sky) number of radio and $\gamma$-ray halos as a
function of flux or surface-brightness thresholds are shown at the top and
bottom panels of Figure 2.  We find that $\sim 30$ $\gamma$-ray halos have
a flux exceeding the EGRET detection threshold of $\sim 10^{-11}{\rm
erg~cm^{-2}~s^{-1}}$. Such halos may therefore constitute, as recently
pointed out by Totani \& Kitayama (2000), a significant fraction of the
unidentified extra-Galactic EGRET sources ($\sim60$ sources over all sky;
see \"{O}zel \& Thompson 1996).  
Note, however, that since the angular extension
of the brightest halos is large, a careful analysis is required to
determine the detectability of such halos by the source search analysis
which was applied to the EGRET data.

\section{Conclusions}

The production of a fluctuating synchrotron background by strong
intergalactic shocks is a natural consequence of structure formation in the
Universe.  The brightest emission originates from the virialization shocks
on Mpc scales around newly formed, massive X-ray clusters. We estimate that
the combined emission from all clusters produces fluctuations in the
microwave background temperature of order $40 \mu{\rm K}
(\xi_B/0.01)(\xi_e/0.05)(\nu/10{\rm GHz})^{-3}$ on sub-degree scales
(Fig. 1), where $\xi_B\sim 0.01$ is the value required to explain the
brightest radio halos of nearby clusters and $\xi_e\sim0.05$ accounts for
the $\gamma$-ray background.  The foreground synchrotron fluctuations might
be comparable to the anisotropy signals detected by existing low-frequency
CMB experiments, and can be easily isolated through multi-frequency
observations. Polarization anisotropy experiments could then constrain the
coherence length of the intergalactic magnetic field.

Our model predicts a fluctuation amplitude $\ga 40\%$ in the $\gamma$-ray
background intensity on sub-degree scale (Fig. 1), and the existence of
extended, $\ga 1\arcdeg$, $\gamma$-ray halos, associated with newly formed
massive clusters (Fig. 2). On scales larger than a degree the fluctuation
amplitude declines and is well below the anisotropy limits from EGRET (see
Fig. 5 in Sreekumar et al. 1998), although somewhat higher than expected
from an analogy to the X-ray background (Loeb \& Waxman 2000).
Detection of the predicted signals will provide a calibration of the
uncertain model parameter $\xi_e$.  The high-energy maps required to detect
the predicted anisotropy signal will be made between 20 MeV and 300 GeV by
the GLAST instrument\footnote {See http://glast.gsfc.nasa.gov/ for more
details.}
(planned for launch in 2005), which is expected to be more sensitive than
EGRET by an order-of-magnitude (\cite{Bloom96}).  The predicted
$\gamma$-ray halos may constitute a significant fraction of the
unidentified extra-Galactic EGRET sources (see also Totani \& Kitayama
2000).  However, since the angular extension of the brightest halos is
large, a more careful analysis is required to assess the detectability of
such halos by EGRET.

A future, dedicated, all-sky anisotropy experiment, operating at several
frequencies below 10 GHz, would be able to map the fluctuations in the
intergalactic synchrotron background.  The resulting synchrotron map could
then be cross-correlated with full-sky maps at hard X-ray or $\gamma$-ray
energies to confirm its cosmic origin. Identification of the synchrotron
fluctuations together with their counterpart inverse-Compton emission of
hard X-rays or $\gamma$-rays by the same population of shock-accelerated
electrons, can be used to empirically determine the strength and spatial
distribution of the intergalactic magnetic fields.  Similarly, the
correlation between radio and $\gamma$-ray halos may be detectable around
individual X-ray clusters.  Strong radio halos could be the best indicators
for bright $\gamma$-ray clusters, which would provide the first obvious
targets for GLAST.

For the sake of simplicity, our model associated the intergalactic shocks
with smooth spherical accretion of gas onto clusters, while in reality they
result from asymmetric mergers as well as from converging flows in large
scale sheets and filaments. The more realistic emission from these complex
geometries can be best modeled through detailed hydrodynamic simulations
(Miniati et al. 2000; Keshet et al. 2000). However, we note that mergers of
comparable mass clusters would tend to produce only mild shocks due to the
prior heating of the shocked gas, and hence result in negligible
non-thermal emission due to the steep power-law slope of the accelerated
electrons.

\acknowledgements
This work was supported in part by grants from the Israel-US BSF 
(BSF-9800343) and NSF
(AST-9900877).  EW thanks the Harvard-Smithsonian Center for Astrophysics
for its kind hospitality during the course of this work. EW is the
incumbent of the Beracha foundation career development chair.

\newpage

\begin{figure}[t]
\centerline{\psfig{figure=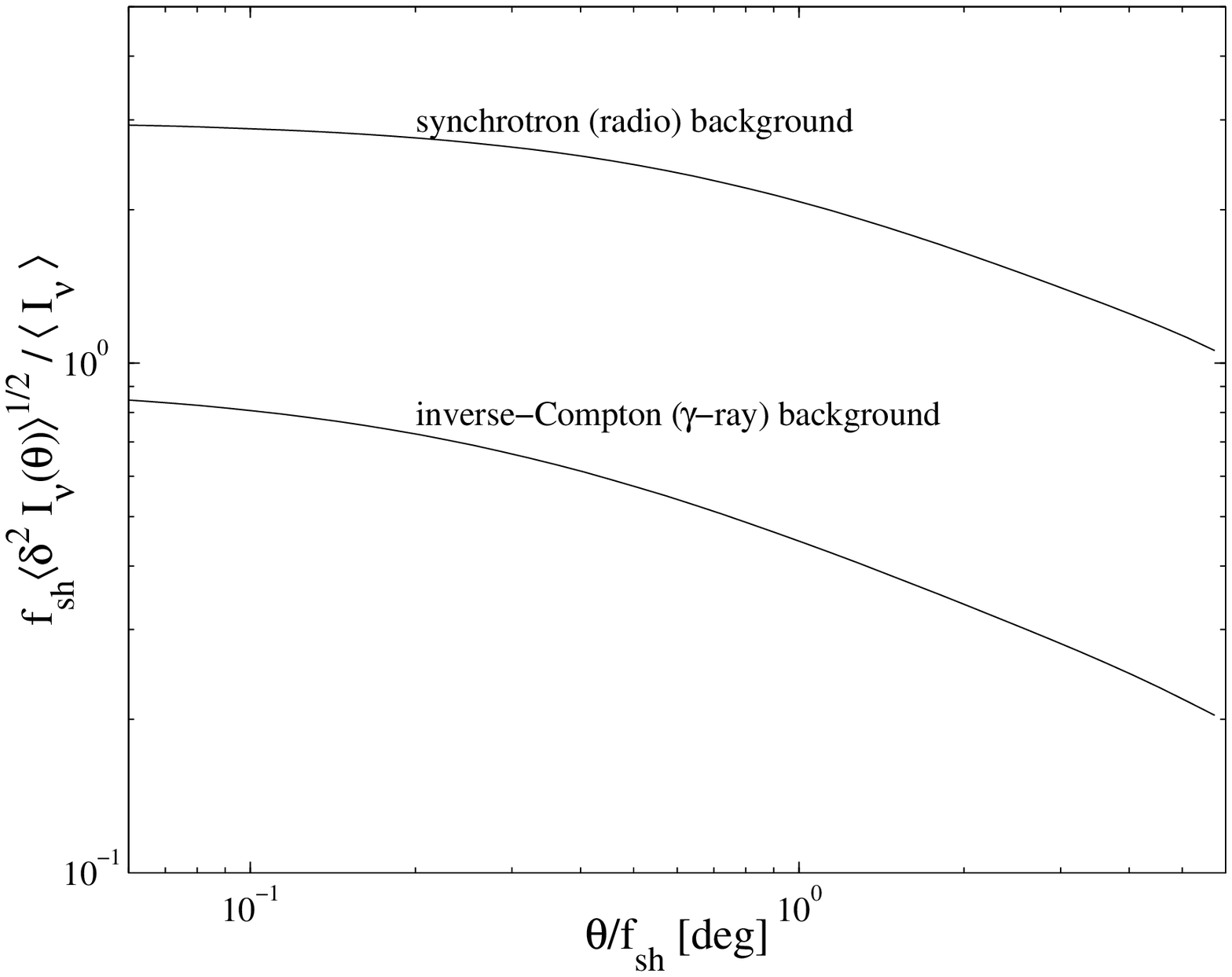,width=6in}}
\caption{ Fractional intensity fluctuations, $(\langle
I_\nu(0)I_\nu(\theta)\rangle-\langle I_\nu\rangle^2)^{1/2}/ \langle I_\nu
\rangle$, in synchrotron (radio) and inverse-Compton ($\gamma$-ray)
background flux. The dimensionless coefficient, $f_{\rm sh}$, is of order
unity [see definition in Eq. (7)].
The ratio of synchrotron
to CMB intensity is $\langle I^{\rm syn}_\nu\rangle/I^{\rm
CMB}_\nu=6\times10^{-6} (f_T/f_{\rm sh}^2)(\xi_B/0.01)(\nu/{10\rm
GHz})^{-3}$, where $f_T$ is a dimensionless coefficient of order unity [see
Eq. (6)], 
and the magnetic energy fraction $\xi_B$ is related to
the magnetic field strength in Eq. (9).
}
\label{fig1}
\end{figure}

\begin{figure}[t]
\centerline{\psfig{figure=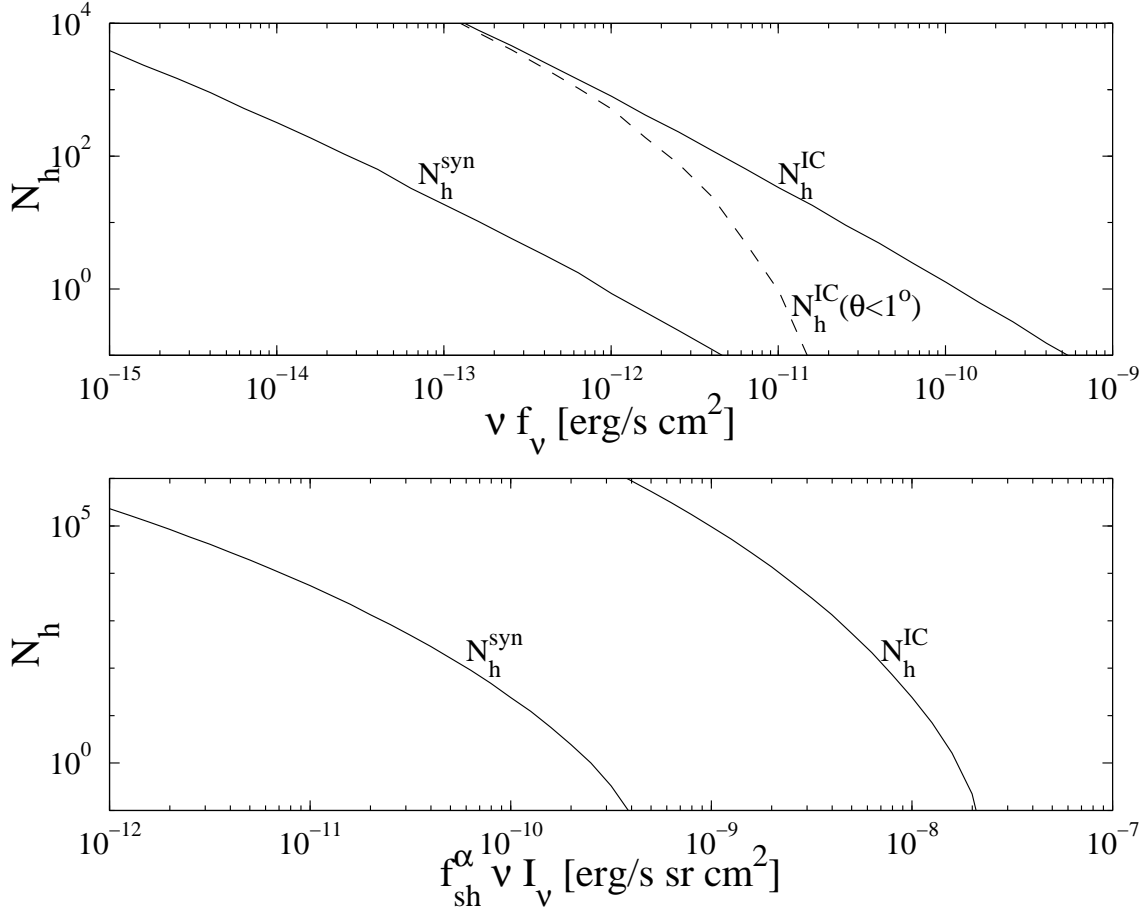,width=6in}}
\caption{ The cumulative all-sky number of inverse-Compton ($\gamma$-ray)
and synchrotron (radio) halos with observed flux (top) or
surface-brightness (bottom) exceeding a value of $\nu f_\nu$ or $\nu
I_\nu$, respectively. The normalization of the inverse-Compton halo curve
is fixed by the requirement that the integrated emission would produce the
observed $\gamma$-ray background. The dashed line represents the number of
$\gamma$- ray halos with angular radius smaller than 1$\arcdeg$.  The
synchrotron halo flux is proportional to $\xi_B$, and the curve shown in the
figure corresponds to $\xi_B=0.01$.  The dimensionless coefficient, $f_{\rm
sh}$, is of order unity [see definition in Eq. (7)]; 
the index
$\alpha=2$ for the inverse-Compton emission, and $\alpha=4$ for the
synchrotron emission.  }
\label{fig2}
\end{figure}

\end{document}